\documentclass[doublecol]{epl2} 
\usepackage{amsbsy}
\usepackage{amstext}
\usepackage{amssymb}
\usepackage{amsmath}

\usepackage[percent]{overpic}

\title{Quantum Flux and Reverse Engineering of Quantum Wavefunctions}
\shorttitle{Quantum Flux and Reverse Engineering of Quantum Wavefunctions} 

\author{D. J. Mason\inst{1} \and M. F. Borunda\inst{2,1} \and E. J. Heller\inst{1,3}}
\shortauthor{D. J. Mason \etal}

\institute{                    
  \inst{1} Department of Physics, Harvard University, Cambridge, Massachusetts
02138, USA\\
  \inst{2} Department of Physics, Oklahoma State University, Stillwater, Oklahoma
74078, USA\\
  \inst{3} Department of Chemistry and Chemical Biology, Harvard University,
Cambridge, Massachusetts 02138, USA
}

\abstract{
An interpretation of the probability flux is given, based on a derivation
of its eigenstates and relating them to coherent state projections
on a quantum wavefunction. An extended definition of the flux operator
is obtained using coherent states. We present a ``processed Husimi''
representation, which makes decisions using many Husimi projections
at each location. The processed Husimi representation reverse engineers
or deconstructs the wavefunction, yielding the underlying classical
ray structure. Our approach makes possible interpreting the dynamics
of systems where the probability flux is uniformly zero or strongly
misleading. The new technique is demonstrated by the calculation of
particle flow maps of the classical dynamics underlying a quantum
wavefunction
\revision{in simple model systems such as a circular billiard with and without a magnetic field}.
}

\begin{document}

\global\long\def\braketop#1#2#3{\left\langle #1\vphantom{#2}\vphantom{#3}\left|#2\vphantom{#1}\vphantom{#3}\right|#3\vphantom{#1}\vphantom{#2}\right\rangle }
\global\long\def\braket#1#2{\left\langle \left.#1\vphantom{#2}\right|#2\right\rangle }
\global\long\def\bra#1{\left\langle #1\right|}
\global\long\def\ket#1{\left|#1\right\rangle }
\global\long\def\ketbra#1#2{\left|#1\vphantom{#2}\right\rangle \left\langle \vphantom{#1}#2\right|}
\global\long\def\figlab#1{(#1)}
\global\long\def\vec#1{\mathbf{#1}}
\global\long\def\coherent#1#2#3{#1,#2,#3}

\maketitle

\section{Introduction}
The probability flux, or probability current, is introduced in quantum
mechanics textbooks as a deterministic operator that can be calculated,
but its connection to experiment is often left to the reader's imagination.
The flux operator, whose expectation over the wavefunction gives the
traditional flux $\vec j\left(\vec r,\vec p\right)$, is defined as
\begin{equation}
\hat{\mathbf{j}}_{\mathbf{r}}=\frac{1}{2m}\left(\ketbra{\mathbf{r}}{\mathbf{r}}\hat{\vec p}+\hat{\vec p}\ketbra{\vec r}{\vec r}\right).\label{eq:Flux-Operator}
\end{equation}

Here $\vec r$ and $\vec p$ indicate the position and momentum of
a particle while $m$ is the mass. The concept of ``flux at a point''
seems paradoxical because we say something about momentum while also
knowing position precisely. This raises the question: Can the flux
even be measured? 

On the other hand, probability flux vanishes on stationary states
for systems with time-reversal symmetry. This is a shame, since strong
semiclassical connections between trajectory flow and quantum eigenstates
lie completely hidden in the universal value of 0 for the flux. This
letter addresses the problem of the underlying dynamics encoded in
stationary states by extending the definition of the flux to coherent
state projections and using a ``processed Husimi''
representation which makes decisions using many Husimi projections
at each location and spits out classical rays. Not only does our approach
resolve the measurement problem for the flux, we also use it as a
technique for extracting semiclassical paths from a quantum wavefunction
even when the flux is zero. 

Several discussions connecting the flux to experimental measurement
exist in the literature \cite{measurement3,Measurement-schrod,measurement2};
we begin our argument instead by identifying the eigenstates of the
flux operator \cite{Park_1988,Garashchuk_2009,Seideman_1991},
and present a physical interpretation. The flux eigenstates had previously
been studied in the context of thermal rates \cite{Park_1988,Seideman_1991}
and of wavepackets\cite{Garashchuk_2009}. By providing a clear derivation
and a novel application of the flux eigenstates, we provide a perspective
of how these techniques can fit into the broader context of wavefunction
analysis.

\section{Eigenstates of the flux}
We begin by replacing the Dirac basis implicit in Eq.~\ref{eq:Flux-Operator}
with the Gaussian basis defined as 
\begin{equation}
\braket{\vec r}{\vec r_{0},\sigma}=N_{\sigma}^{d/2}e^{-\left(\vec r-\vec r_{0}\right)^{2}/4\sigma^{2}},\label{eq:Gaussian}
\end{equation}
where $d$ is the number of dimensions in the system and $N_{\sigma}=\left(\sigma\sqrt{2\pi}\right)^{-1}$
is a normalization constant. There is certainly no loss of generality
here; the $\sigma\rightarrow0$ Dirac delta basis limit is always
just a step away. Instead, there is new capability introduced, as
we show below. The Gaussian basis over all centers is overcomplete,
so nothing is left hidden from the analysis we develop now. An alternate
derivation using only coherent states confirms the validity of our
results. Below we are led to a harmonic oscillator basis at each location.
In the Gaussian basis, the modified flux operator is 
\begin{align*}
\hat{\vec j}_{\vec r_{0},\sigma}=\frac{1}{2m}\left(\ketbra{\vec r_{0},\sigma}{\vec r_{0},\sigma}\hat{\vec p}+\hat{\vec p}\ketbra{\vec r_{0},\sigma}{\vec r_{0},\sigma}\right). \nonumber
\end{align*}
The eigenstates, projected onto each orthogonal spatial dimension
$i$, are obtained using the eigenvalue equation 
\begin{equation}
\hat{j}_{\vec r_{0},\sigma,i}\ket{\lambda_{\sigma,i}}=\lambda_{\sigma,i}\ket{\lambda_{\sigma,i}},\label{eq:EigenEq}
\end{equation}
which has a solution of the form 
\begin{align*}
\ket{\lambda_{\sigma,i}}=\ket{\vec r_{0},\sigma}+a\hat{p}_{i}\ket{\vec r_{0},\sigma}.
\end{align*}
Using the following two equations 
\begin{align*}
\braketop{\vec r}{\hat{\vec p}}{\vec r_{0},\sigma}=i\hbar\sigma^{-2}\left(\vec r-\vec r_{0}\right)e^{-\left(\vec r-\vec r_{0}\right)^{2}/4\sigma^{2}} 
\end{align*}
and $\braketop{\vec r_{0},\sigma}{\hat{\vec p}}{\vec r_{0,}\sigma}=0,$
we can write 
\begin{equation}
\hat{j}_{\vec r_{0},\sigma,i}\ket{\lambda_{\sigma,i}}=\frac{1}{2m}\left(a\left\langle \hat{p}_{i}^{2}\right\rangle _{\sigma}\ket{\vec r_{0},\sigma}+\hat{p}_{i}\ket{\vec r_{0},\sigma}\right).\label{eq:Explicit-EigenEq}
\end{equation}
Finding the conditions on $\lambda_{\sigma,i}$ that allow Eq.~\ref{eq:Explicit-EigenEq}
to be written in the form of Eq.~\ref{eq:EigenEq}, we obtain 
\begin{align*}
\lambda_{\sigma,i}=\frac{a}{2m}\left\langle \hat{p}_{i}^{2}\right\rangle _{\sigma};\lambda_{\sigma,i}=\frac{1}{2ma}.
\end{align*}

Since $\left\langle \hat{p}_{i}^{2}\right\rangle _{\sigma}=\frac{\hbar^{2}}{4\sigma^{2}}$,
we find the value of $a=\pm\frac{2\sigma}{\hbar}$ from which we obtain
the two eigenvalues 
\begin{equation}
\lambda_{\sigma,i,\pm}=\pm\frac{\hbar}{4m\sigma}.\label{eq:Flux-eigenvalue}
\end{equation}
The eigenstates take the form 
\begin{equation}
\braket{\vec r}{\lambda_{\sigma,i,\pm}}=\braket{\vec r}{\vec r_{0},\sigma}\pm\frac{i}{\sigma}\vec e_{i}\cdot\left(\vec r-\vec r_{0}\right)\braket{\vec r}{\vec r_{0},\sigma},\label{eq:Flux-Eigenstate}
\end{equation}
where $\vec e_{i}$ is the unit vector along spatial direction $i$.
Eq.~\ref{eq:Flux-Eigenstate} is a linear combination of two functions:
the Gaussian (Eq.~\ref{eq:Gaussian}) and its derivative. Projection
of a wavefunction onto the first term can be interpreted as measuring
its probability amplitude at point $\vec r_{0}$, and projection onto
second term as measuring its derivative along the $i^{th}$
direction at the point $\vec r_{0}$.

\section{Expectation value of the flux operator} 
To determine the expectation value of the flux operator, we begin
by labeling the excited states of the harmonic oscillator at position
$\vec r_{0}$ oriented along the $i^{th}$ direction 
\begin{eqnarray*}
\braket{\vec r}0 & = & \braket{\vec r}{\vec r_{0},\sigma}\\
\braket{\vec r}1 & = & \frac{\vec e_{i}\cdot(\vec r-\vec r_{0})}{\sigma}\braket{\vec r}{\vec r_{0},\sigma}\\
\braket{\vec r}2 & = & \sqrt{\frac{1}{2}}\left(\frac{\left(\vec e_{i}\cdot(\vec r-\vec r_{0})\right)^{2}}{\sigma^{2}}-1\right)\braket{\vec r}{\vec r_{0},\sigma}\mbox{; etc.}
\end{eqnarray*}
These states form a complete set in which the flux operator can be
explicitly expressed as the Hermitian matrix
\begin{align*}
\hat{j}_{\vec r_{0},\sigma,i}=\left(\begin{array}{ccccc}
0 & +i\lambda & 0 & \cdots & 0\\
-i\lambda & 0 & 0 & \cdots & 0\\
0 & 0 & 0 & \cdots & 0\\
\vdots & \vdots & \vdots & \ddots & \vdots\\
0 & 0 & 0 & \cdots & 0
\end{array}\right)
\end{align*}
where $\lambda=\lambda_{\sigma,i,+}=\frac{\hbar}{4m\sigma}$. There
are additional sets of harmonic oscillator states centered at points
other than $\vec r_{0}$ also with zero components in the flux matrix.

The complete set of eigenstates $\ket{\lambda_{1}},\ket{\lambda_{2}},\ket{\lambda_{3}},\dots$
of the flux operator expressed in terms of excited states of the harmonic
oscillator are 
\begin{align*}
\ket{\lambda_{1}},\ket{\lambda_{2}},\ket{\lambda_{3}},\dots=\left(\begin{array}{c}
+1\\
-i\\
0\\
\vdots
\end{array}\right),\left(\begin{array}{c}
+1\\
+i\\
0\\
\vdots
\end{array}\right),\left(\begin{array}{c}
0\\
0\\
1\\
\vdots
\end{array}\right),\cdots
\end{align*}
with eigenvalues $-\lambda$, $\lambda$, and $0$. Measurement by
the flux operator collapses the wavefunction onto one of these eigenstates,
the infinite majority of which are in the degenerate zero-eigenvalue
subspace spanning all excited states of the harmonic oscillator above
$\ket 1$. Only the first two eigenstates yield non-zero flux values.
In the $\sigma\rightarrow0$ limit, the eigenvalues of these two states
tend towards positive and negative infinity. 

When expanding the flux expectation value, we can use the complete
eigenbasis to show that 

\begin{eqnarray}
\braketop{\psi}{\hat{j}_{\vec r_{0},\sigma,i}}{\psi} & = & \braketop{\psi}{\hat{j}_{\vec r_{0},\sigma,i}\sum_{i=1}^{\infty}\ketbra{\vphantom{\sum_{i=1}^{\infty}}\lambda_{i}}{\lambda_{i}}}{\psi}\nonumber \\
 & = & \lambda\left|\braket{\psi}{\lambda_{1}}\right|^{2}-\lambda\left|\braket{\psi}{\lambda_{2}}\right|^{2}.\label{eq:Flux-Eigenstate-Simplified}
\end{eqnarray}

From Eq.~\ref{eq:Flux-Eigenstate}, it can be shown that some contributions
from $\left|\braket{\psi}0\right|^{2}$ and $\left|\braket{\psi}1\right|^{2}$
cancel themselves due to the opposite sign of the eigenvalues, and
only the cross-term $\braket{\psi}0^{\ast}\braket{\psi}1-\braket{\psi}0\braket{\psi}1^{\ast}$
remains. This form is directly related to the following expression
for the flux at point $\vec r_{0}$ 
\begin{align*}
\vec j_{\vec r_{0}}\left(\Psi(\vec r)\right)=\frac{\hbar}{2mi}\left(\Psi^{*}(\vec r_{0})\mathbf{\nabla}\Psi(\vec r_{0})-\Psi(\vec r_{0})\mathbf{\nabla}\Psi^{*}(\vec r_{0})\right).
\end{align*}
Eq.~\ref{eq:Flux-Eigenstate-Simplified} can be rewritten as 
\begin{eqnarray}
\braketop{\psi}{\hat{j}_{\vec r_{0},\sigma,i}}{\psi} & = & \frac{i\hbar}{4m\sigma^{2}}[\braketop{\psi}{\vec e_{i}\cdot\left(\vec r-\vec r_{0}\right)}{\vec r_{0},\sigma}\braket{\psi}{\vec r_{0},\sigma}^{\ast}\nonumber \\
 &  & -\braketop{\psi}{\vec e_{i}\cdot\left(\vec r-\vec r_{0}\right)}{\vec r_{0},\sigma}^{\ast}\braket{\psi}{\vec r_{0},\sigma}].\label{eq:Flux-Expectation}
\end{eqnarray}

The traditional flux operator corresponds to the limit $\sigma\rightarrow0$,
at which point the two terms in Eq.~\ref{eq:Flux-Eigenstate} become
the delta function and its derivative, while the flux values of the
first two eigenstates become 
\begin{align*}
\lim_{\sigma\rightarrow0^{+}}\lambda_{\sigma,i,\pm}=\pm\infty.
\end{align*}
In addition, there are an infinite number of other eigenstates with
zero eigenvalues. The infinite value for the two non-zero eigenstates
is corraborated by the observation of Park and Light \cite{Park_1988},
however, in their work they reach this conclusion in the limit of
an infinitely large basis \cite{Seideman_1991}.

\section{Flux measurement}
A single application of the flux operator at a particular point in
space almost always results in zero, but occasionally in an immensely
positive or negative value. It is thus necessary to perform the averaging
over an infinite number of measurements to obtain an expression equivalent
to the textbook flux. 

The prefactors before the Gaussian states in Eq.~\ref{eq:Flux-Eigenstate}
are related to the Taylor expansion 
\begin{align*}
e^{\pm\frac{i}{\sigma}\vec e_{i}\cdot(\vec r-\vec r_{0})}\approx1\pm\frac{i}{\sigma}\vec e_{i}\cdot(\vec r-\vec r_{0}).
\end{align*}

This suggests a connection between the flux eigenstates and the coherent
state, defined as
\begin{equation}
\braket{\vec r}{\vec r_{0},\vec k_{0},\sigma}=N_{\sigma}^{d/2}e^{-\left(\vec r-\vec r_{0}\right)^{2}/4\sigma^{2}+i\vec k_{0}\cdot\vec r},\label{eq:Coherent-State}
\end{equation}
which is a Gaussian envelope over the plane wave $e^{i\vec k_{0}\cdot\vec r}$.
Observing that the phase $e^{i\vec k_{0}\cdot\vec r_{0}}$ is arbitrary,
we Taylor expand in the limit of $\vec k_{0}\sigma\ll1$ to produce

\begin{eqnarray}
\braket{\vec r}{\vec r_{0},\vec k_{0},\sigma} & \approx & N_{\sigma}^{d/2}e^{-\left(\vec r-\vec r_{0}\right)^{2}/4\sigma^{2}}\left(1+i\vec k_{0}\cdot\left(\vec r-\vec r_{0}\right)\right)\nonumber \\
 & \approx & \braket{\vec r}{\vec r_{0},\sigma}+i\vec k_{0}\cdot\left(\vec r-\vec r_{0}\right)\braket{\vec r}{\vec r_{0},\sigma}.\label{eq:Taylor-Expand}
\end{eqnarray}
Note that the dispersion relation for the free-particle continuum
is a circle with radius $k_{0}=\frac{\sqrt{2mE}}{\hbar}$, which does
not depend on the orientation of $\vec k_{0}$. The second terms in
Eq.~\ref{eq:Taylor-Expand} and Eq.~\ref{eq:Flux-Eigenstate} are
proportional to each other when $\vec k_{0}$ points along the $i^{th}$
direction. This similarity allows us to relate the flux expectation
value from Eqs.~\ref{eq:Flux-Eigenstate-Simplified} and \ref{eq:Flux-Expectation}
to coherent state projections as

\begin{eqnarray}
\lim_{\sigma k_{0}\rightarrow0}\braketop{\psi}{\hat{j}_{\vec r_{0},\sigma,i}}{\psi} & = & \frac{\hbar k_{0}}{4m\sigma^{2}}[\left|\braket{\psi}{\vec r_{0},k_{0}\vec e_{i},\sigma}\right|^{2}\nonumber \\
 &  & -\left|\braket{\psi}{\vec r_{0},-k_{0}\vec e_{i},\sigma}\right|^{2}],\label{eq:Flux-Hus-Corr}
\end{eqnarray}
where the traditional flux \emph{vector} is constructed from the components
in each direction.

\section{Processed Husimi representation}
The representation of quantum mechanical systems in terms of phase-space
distribution functions, such as Wigner's quasi-probability distribution
function\cite{PhysRev.40.749,Hillery1984121} and the Husimi distribution
function\cite{Husimi}, allows expressing quantum mechanical results
into a form that resembles classical mechanics\cite{hellerleshouches,Husimi-map-old,Husimi-map-old2}.
The standard Husimi projection technique consists of an idealized
measurement of position and momentum consistent with the uncertainty
principle. The Husimi distribution consists of the expectation value
of the projector onto a coherent state and is a phase space probability
density.

Based on the coherent state projection of the flux operator, we can
now develop a deeper intuition. The processed Husimi technique uses
coherent states to produce maps of the flux that display information
about the local phase space distribution of the wavefunction beginning
in real space. As outlined below, the current flow maps adhere to
the uncertainty principle constraints. Applying the processed Husimi
representation to a quantum wavefunction can reverse engineer or deconstruct
it, yielding the underlying classical rays even when those rays cross
in several directions, as we demonstrate below.

Using the uncertainty relation $\Delta x\propto1/\Delta k\propto\sigma$
and setting $\sigma\rightarrow0$ results in coherent state measurements
with infinite uncertainty in $k$-space and zero uncertainty in real
space. The traditional flux therefore operates in the limit of infinite
momentum uncertainty. For small $\sigma$, due to the large momentum
uncertainty, coherent state projections merely reproduce the probability
amplitude $\left|\psi(\vec r)\right|^{2}$ in all directions of $\vec k_{0}$.
\textit{The flux emerges as a small residual which can be retrieved
by summing each coherent state projection weighted by $\vec k_{0}$,}
such that 
\begin{equation}
\braketop{\psi}{\hat{\vec j}_{\vec r_{0},\sigma}}{\psi}\approx\int\vec k_{0}\left|\braket{\psi}{\vec r_{0},\vec k_{0},\sigma}\right|^{2}d^{d}k_{0}.\label{eq:Husimi-Vector}
\end{equation}
Note that in the limit $\sigma\rightarrow0$, the contributing points
in the $k$-space integral reduce to just the orthogonal directions.
The absence of flux in time-reversal symmetric states can be interpreted
as the mutual cancellation of coherent state projections along each
direction in $k$-space.

For larger $\sigma$, reduced momentum uncertainty allows for substantial
variation in the coherent state projections between different directions
of $\vec k_{0}$. In this regime, we can use coherent states (Husimi
projections) to produce a map of the local phase space of a wavefunction.
By taking snapshots of the phase space at many points across a system
for larger $\sigma$, we can process the result to produce a semiclassical
map showing the dominant classical paths contributing to a given wavefunction.
Thus, the term \textit{processed Husimi} for these visualizations.
Like the traditional flux map, processed Husimi flows can be integrated
over lines and surfaces to reveal net current flow.

\begin{figure}
\onefigure[width=0.9\columnwidth]{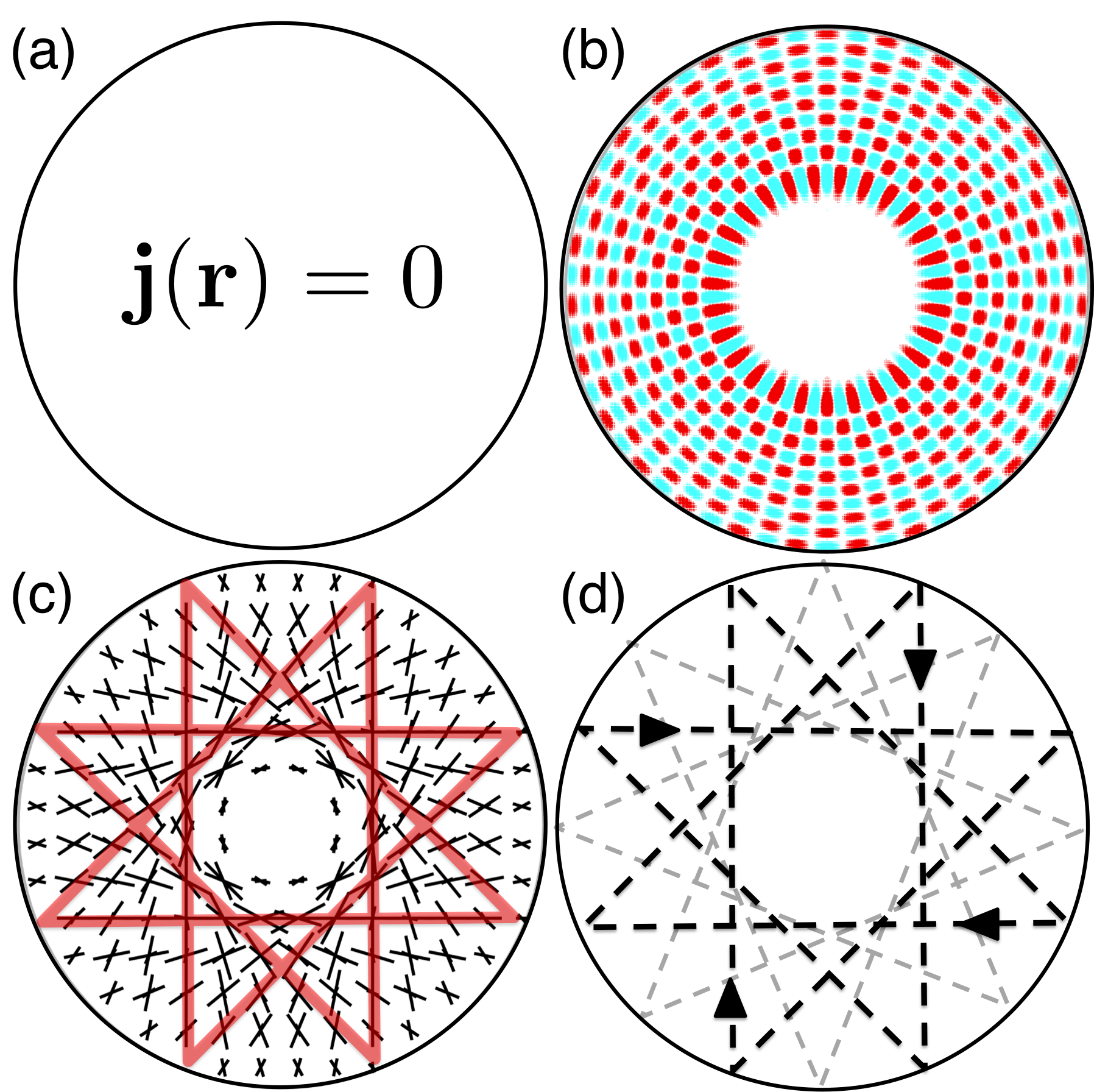}
\caption{States of the circular billiard exhibit defined
angular momentum and rotational symmetry. The flux (a) is uniformly
zero due to time-reversal symmetry. The wavefunction (b), processed
Husimi flow (c), and corresponding classical paths (d) are shown for
an eigenstate of the circular well in which the angular and radial
components of the wavefunction are nearly equal. The classical path
in (d) is reproduced in (c) to highlight its correspondence with the
processed Husimi flow.}\label{fig:Ang-Mom}
\end{figure}

Although the calculations reported here are for very simple models,
we believe that the technique is generic and can be applied to virtually
any quantum wavefunction and model amenable to semiclassical analysis\cite{Mason-Husimi-Continuous,Mason-Husimi-Lattices,Mason-Husimi-Graphene}.
We begin by demonstrating the processed Husimi technique on a circular
billiard, which, due to time-reversal symmetry, has zero flux. The
Schr\"{o}dinger equation for this system can be written in radial form
as 
\begin{align*}
\frac{d^{2}R(r)}{dr^{2}}+\frac{1}{r}\frac{dR(r)}{dr}+\left(k^{2}-\frac{m^{2}}{r^{2}}\right)R(r)=0.
\end{align*}
Solutions to this equation are simultaneous eigenstates of energy
and angular momentum, and thus possess the good quantum numbers $n$
(number of nodes in the radial direction) and $m$ (number of angular
nodes). Fig.~\ref{fig:Ang-Mom}b shows one such state with $n\approx m$,
which corresponds to classical paths that bounce off the boundary
at a consistent $30^{\circ}$ angle (see schematic in Fig.~\ref{fig:Ang-Mom}d). 

Because the flux is uniformly zero, it is no help in understanding
the dynamics of this system. In Fig.~\ref{fig:Ang-Mom}c we present
the processed Husimi flow of this wavefunction by sampling coherent
state projections on a regularly-spaced grid across the system. With
the solid line, we show that these projections align perfectly with
one of the classical paths indicated in Fig.~\ref{fig:Ang-Mom}d.
In addition, each point in the Husimi map contains an additional set
of Husimi vectors, which do not align with the path. Given that any
state of a circularly symmetric system must correspond with infinitely
many classical trajectories related by rotation, the processed Husimi
at a particular point must reflect all rotated paths that intersect
there. We indicate one classical path and its rotated version in grey
in Fig.~\ref{fig:Ang-Mom}d. The ``cross-hatching'' pattern arises
because two rotated classical paths intersect at any point, explaining
the similar cross-hatching nodal patterns in the wavefunction.

Magnetic systems can also be analyzed using processed Husimi projections,
thus shining light on recent work examining flux vortices in quantum
dots\cite{Vortex1,Vortex2}. Time-reversal symmetry in the circular
billiard is broken by the magnetic field. To properly represent these
states, both the momentum operator in the flux operator (Eq.~\ref{eq:Flux-Operator})
and the momentum term $i\vec k_{0}\cdot\vec r_{0}$ in the coherent
state (Eq.~\ref{eq:Coherent-State}) must be modified to reflect
the canonical transformation $\vec p\rightarrow\vec p-q\vec A/c$,
where $\vec A$ is the magnetic potential.

\begin{figure}
\onefigure[width=0.9\columnwidth]{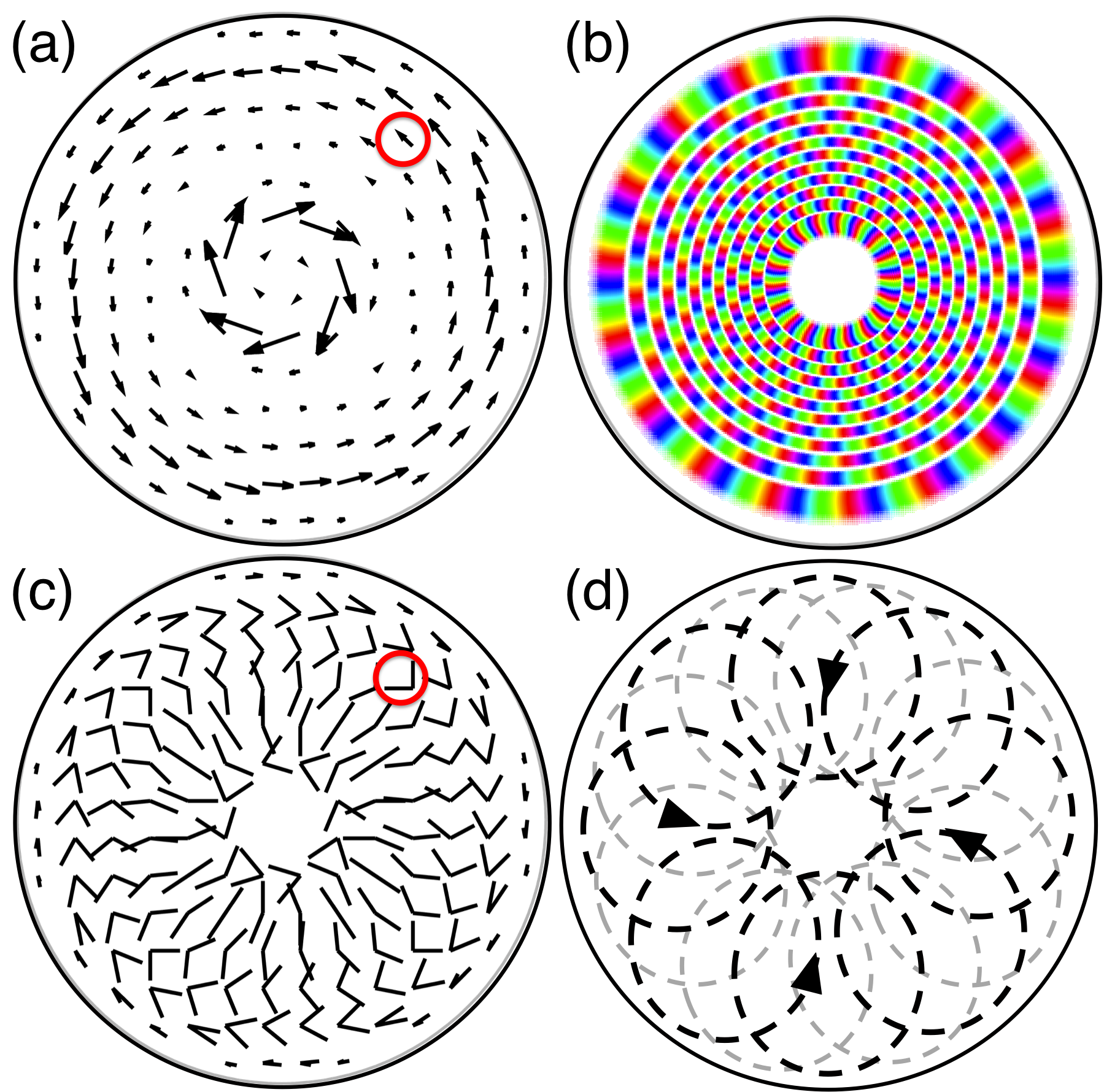}

\caption{\label{fig:Mag-Field}States of the circular billiard with an applied
magnetic field correspond to circular classical paths with a unique
cyclotron radius. The flux (a), wavefunction (b), processed Husimi
flow (c), and corresponding classical paths (d) are shown for an eigenstate
of the circular well with strong magnetic field perpendicular to the
plane. The cyclotron radius for this state is approximately one-third
of the system radius, which is corroborated in the processed Husimi
flow, but not the flux.}
\end{figure}

\begin{figure}
\onefigure[width=0.9\columnwidth]{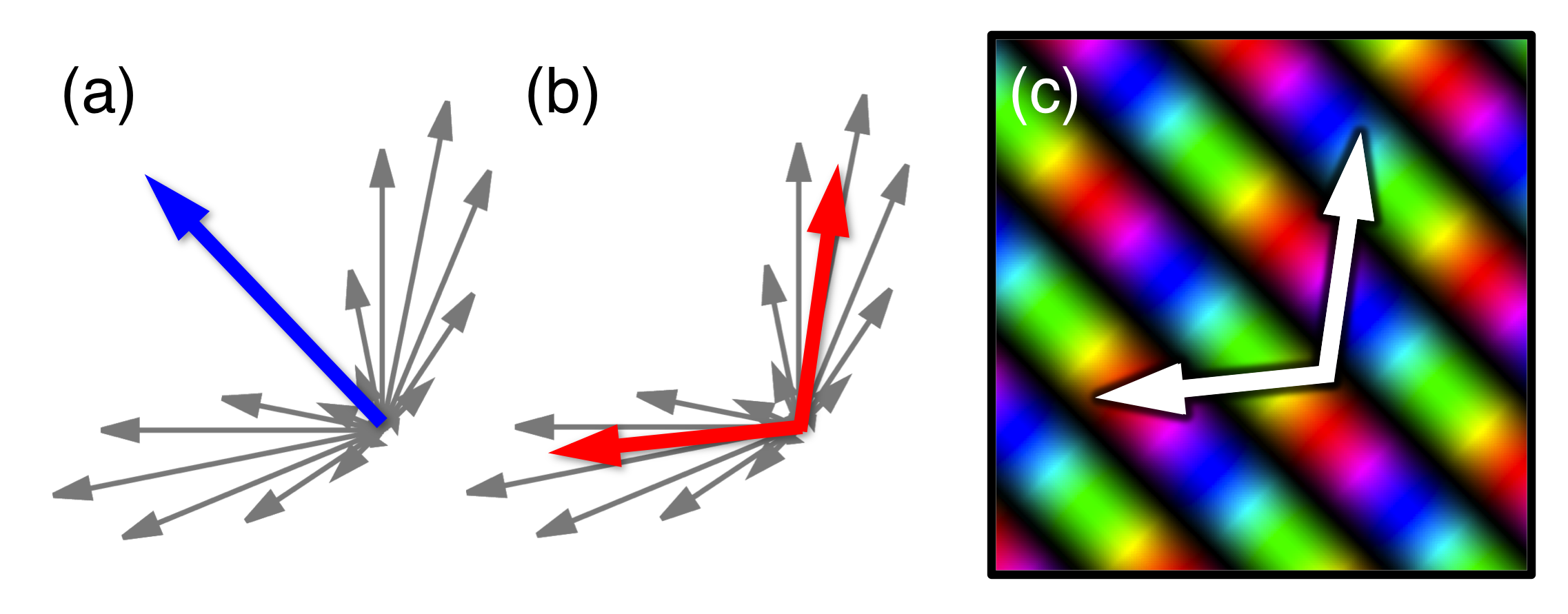}

\caption{\label{fig:Flux-Multi-Modal}Processed Husimi reconstruction of the
classical trajectories. Coherent state projection vectors for 32 equally-space
points in $k$-space are shown in grey (a-b) for the double plane
waves (c) defined in Eq.~\ref{eq:Psi-3}. Because of momentum uncertainty,
there is spread in the vectors. The flux operator (a, blue) averages
over the vectors of the coherent state projections, while the processed
Husimi (b, red) recovers both underlying directions. The wavefunction
in (c) is representative of the areas circled in Fig.~\ref{fig:Mag-Field}.}
\end{figure}

For strong enough magnetic fields, classical trajectories are circular
with radii corresponding to the cyclotron radius. However, the dynamics
due to the cyclotron radius are not obvious from the wavefunction
in Fig.~\ref{fig:Mag-Field}b. Worse still, the flux map in Fig.~\ref{fig:Mag-Field}a
(obtained by sampling at regularly-spaced points in a grid similar
to the one used for the processed Husimi flow) appears to indicate
\emph{two} circular paths but with strongly different radii.

The processed Husimi flow in Fig.~\ref{fig:Mag-Field}c dramatically
clarifies the dynamics by indicating classical paths consistent with
a cyclotron radius approximately one-third of the radius of the system.
The cyclotron orbits are offset from the center of the system; due
to circular symmetry, there exist infinitely many such cyclotron orbits
rotated around the system center. We show a subset of these orbits
in the schematic in Fig.~\ref{fig:Mag-Field}d, which closely parallels
the processed Husimi flow and exhibits the correct cyclotron radius.

Why does the flux map fail to show the dominant classical paths? In
Fig.~\ref{fig:Flux-Multi-Modal}, we provide magnified views from
the full set of coherent state projections, the flux map, and the
processed Husimi flow corresponding to the circles in Figs.~\ref{fig:Mag-Field}.
We can model this point in the wavefunction according to the pure
momentum state 
\begin{equation}
\Psi\left(\vec r\right)=e^{i\vec k_{1}\cdot\vec r}+e^{i\vec k_{2}\cdot\vec r},\label{eq:Psi-3}
\end{equation}
where $\vec k_{1}$ and $\vec k_{2}$ are indicated by the arrows
in Fig.~\ref{fig:Flux-Multi-Modal}c. Because of the uncertainty
of each coherent state projection, the full set of projections exhibit
a finite spread around the generating wavevectors. The processed Husimi
technique (red) retrieves two independent trajectories at this point,
while the flux (blue) averages over them. The flux map amounts to
summing the Husimi vectors, giving a total flow at each point that
does not always correspond to the semiclassical dynamics of the system.
With the processed Husimi flow, we now have a complementary representation
for revealing the flow structure present in the wavefunction.

Processed Husimi maps also have implications for experiments measured
in a fashion similar to angle-resolved photoemission spectroscopy
(for a review, see \cite{ARPES-Review}). In the ARPES setup, a focused
photon beam on a sample emits electrons from the valence band. The
energy of the photo-emitted electrons incorporates both their bonding
energies, which can be averaged over, and their kinetic energy, which
depends on the angle of the beam with respect to the sample surface. 

The ARPES response function behaves similarly to coherent state projections
with $\vec k_{0}$ proportional to the beam angle. By rotating the
beam angle around the same point of intersection, the response in
different directions provides the momentum distribution of the wavefunction
at that point. Perturbations from the known dispersion relation can
then be inserted into Eq.~\ref{eq:Husimi-Vector} to obtain the flux
expectation value. 

While a narrow beam could measure the flux vector at the intersection
point, it will be difficult to distinguish the occasional large perturbation
measurements from noise. However, wider beams could capture additional
terms from the Taylor expansion of the coherent state in Eq.~\ref{eq:Taylor-Expand},
producing more reliable measurements. Applying the technique at many
points across the sample would then provide the processed Husimi map.

\section{Conclusions}
We have provided a new interpretation of the flux operator from the
perspective of its eigenstates while connecting them to coherent state
projections at the limit of infinitesimal spatial spread. Away from
this limit, we can use coherent state projections to provide a map
of the classical dynamics underlying a quantum wavefunction, permitting
us to describe flow even in stationary states with zero flux. In systems
with magnetic fields, we have shown that the flux maps actually correspond
to the aggregate of such classical flows, which we are able to retrieve
from the processed Husimi projections.

\acknowledgments
This research was conducted with funding from the Department of Energy
Computer Science Graduate Fellowship program under Contract No. DE-FG02-97ER25308.
MFB and EJH were supported by the Department of Energy, office of
basic science (grant DE-FG02-08ER46513).

\end{document}